\documentclass[prl,preprint]{revtex4}
\usepackage{graphicx}
\usepackage{dcolumn}
\usepackage[T1]{fontenc}
\usepackage{amssymb, amsmath,amsfonts}
\usepackage{bm}
\usepackage[usenames]{color}
\usepackage{hyperref}

\definecolor{Gr}{rgb}{0,0.3,0}

\begin{document}

 \title{Misfit layer compounds: a platform for heavily-doped  two-dimensional transition metal dichalcogenides.}
   
\author{Rapha\"el T. Leriche$^1$}
\author{Alexandra Palacio-Morales$^{1,2}$}
\author{Marco Campetella$^1$}
\author{Cesare Tresca$^1$}
\author{Shunsuke Sasaki$^3$}
\author{Christophe Brun$^1$}
\author{Fran\c{c}ois Debontridder$^1$}
\author{Pascal David$^1$}
\author{Imad Arfaoui$^4$}
\author{Ondrej \v{S}ofranko$^{5,6}$}
\author{Tomas Samuely$^5$}
\author{Geoffroy Kremer$^7$}
\author{Claude Monney$^7$}
\author{Thomas Jaouen$^{7,8}$}
\author{Laurent Cario$^3$}
\author{Matteo Calandra$^{1,9,10}$}
\email{m.calandrabuonaura@unitn.it}
\author{Tristan Cren$^1$}
\email{tristan.cren@upmc.fr}

\affiliation{$^1$Institut des NanoSciences de Paris, Sorbonne Universit\'e and CNRS-UMR 7588, 75005 Paris, France}
\affiliation{$^2$Laboratoire de Physique des Solides, Universit\'e Paris-Saclay and CNRS UMR8502, 91405 Orsay, France}
\affiliation{$^3$Institut des Mat\'eriaux Jean Rouxel, Universit\'e de Nantes and CNRS-UMR 6502, 44322 Nantes,  France}
\affiliation{$^4$Monaris, Sorbonne Universit\'e and CNRS-UMR 8233, 75005 Paris, France}
\affiliation{$^5$Centre of Low Temperature Physics, Faculty of Science, P. J. Safarik University, SK-04001 Kosice, Slovakia}
\affiliation{$^6$Centre of Low Temperature Physics, Institute of Experimental Physics, Slovak Academy of Sciences, SK-04001 Kosice, Slovakia}
\affiliation{$^7$D\'epartement de Physique and Fribourg Center for Nanomaterials, Universit\'e de Fribourg, CH-1700 Fribourg, Switzerland}
\affiliation{$^8$Univ Rennes, CNRS, IPR (Institut de Physique de Rennes) - UMR 6251,  F-35000 Rennes, France}
\affiliation{$^9$ Department of Physics, University of Trento, 38123 Provo, Italy}
\affiliation{$^{10}$ Graphene Labs, Fondazione Istituto Italiano di Tecnologia, Via Morego, I-16163 Genova, Italy}

\date{\today}

\maketitle

\begin{bf}
Transition metal dichalcogenides (TMDs) display a rich  variety of instabilities such as spin and charge orders\cite{monceau_electronic_2012, WilsonDisalvo}, Ising superconductivity \cite{yuan_possible_2014, lu_evidence_2015, xi_ising_2016, zhou_ising_2016, hsu_topological_2017} and topological properties \cite{qian_solid_2014, fei_edge_2017}. Their physical properties can be controlled by doping in electric double-layer field-effect transistors (FET). However, 
for the case of single layer NbSe$_2$, FET doping is limited to $\approx 1\times 10^{14}$  cm$^{-2}$ \cite{xi_gate_2016}, while a somewhat larger charge injection can be obtained via  deposition of K atoms.
Here, by performing ARPES, STM, quasiparticle interference measurements, and first principles calculations we show that a misfit compound formed by sandwiching NbSe$_2$ and LaSe layers behaves as a NbSe$_2$ single layer
with a rigid doping of $0.55-0.6$ electrons per Nb atom or $\approx 6\times 10^{14}$  cm$^{-2}$.  Due to this huge doping, the $3\times3$ charge density wave is replaced by a $2\times2$ order with very short coherence length. As a tremendous number of different misfit compounds can be obtained by sandwiching TMDs layers with rock salt or other layers \cite{wiegers_misfit_1992, wiegers_misfit_1996-1}, 
our work paves the way to the exploration of heavily doped 2D TMDs over an unprecedented wide range of doping.
\end{bf}

\section{Introduction}

\begin{figure*}[h!]
\includegraphics[width = 14 cm]{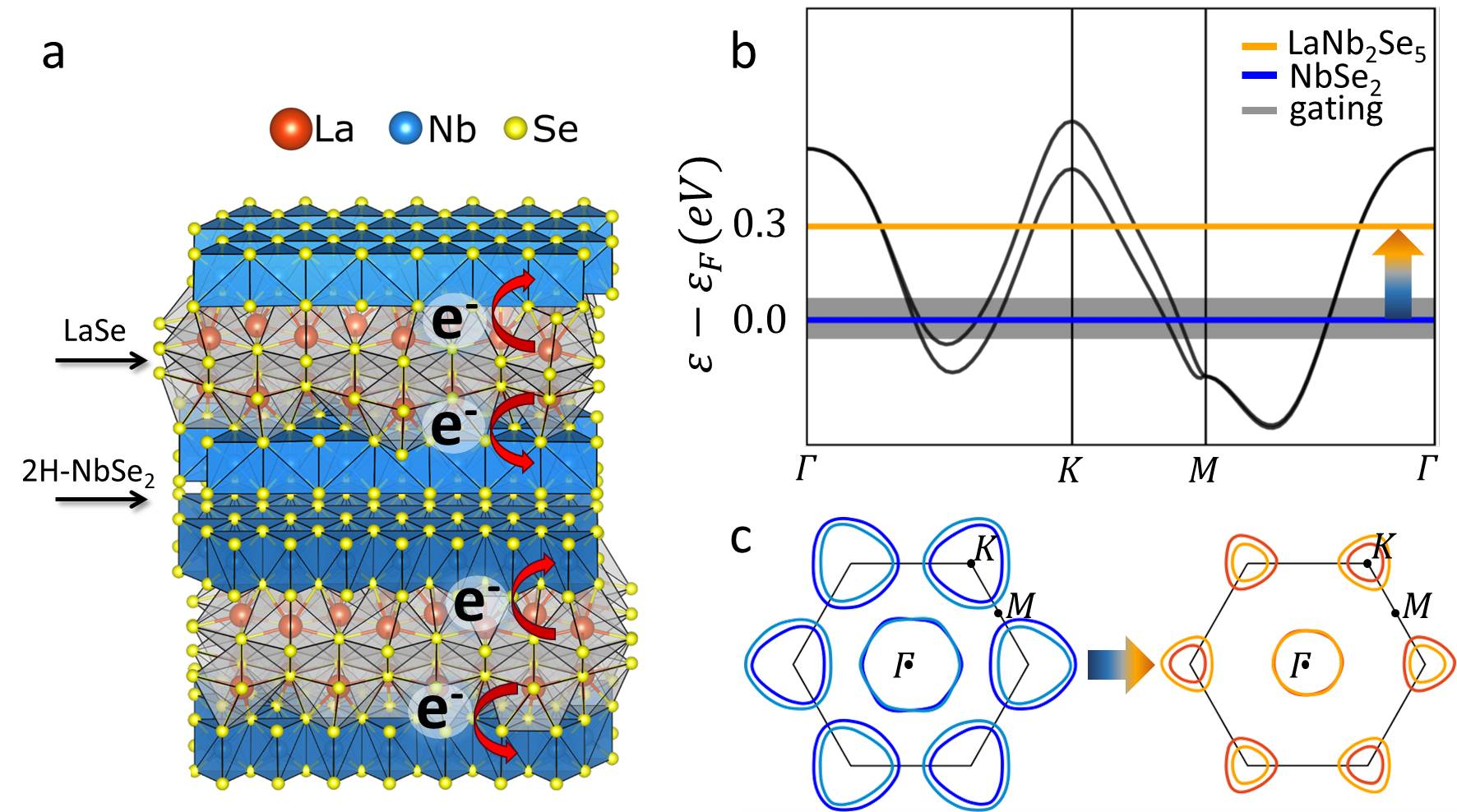}
\caption{(a) crystal structure of (LaSe)$_{1.14}$(NbSe$_2$)$_2$, the material is made of a stacking of LaSe and NbSe$_2$ layers. The LaSe layer is a massive electron donor. The electronic structure of (LaSe)$_{1.14}$(NbSe$_2$)$_2$ is very close to the one of a strongly doped NbSe$_2$ monolayer which energy dispersion curves around the Fermi level are illustrated by black curves in (b). The gray stripe illustrates the range of doping accessible by gating while the yellow line shows the huge doping induced by charge transfer from the LaSe layer. This huge doping results in a shrunk Fermi surface shown in (c). The Fermi surface of an undoped NbSe$_2$ monolayer is shown in blue, the one of (LaSe)$_{1.14}$(NbSe$_2$)$_2$ is shown in yellow. } 
\label{fig1}
\end{figure*}

TMDs are layered materials with strong in-plane bonding and weak, van der Waals-like, coupling between the layers \cite{WilsonDisalvo}. They exhibit a rich physics with many kinds of electronic orders such as charge density wave or superconductivity \cite{monceau_electronic_2012, klemm_layered_2011}. 
In this family, single layer NbSe$_2$ attracted a considerable interest as it presents a charge density wave instability with $3\times3$ periodicity \cite{xi_strongly_2015,ugeda_characterization_2016} and a strong spin-orbit interaction leading to spin-momentum locking in the out of plane direction generating fully spin up and spin down polarized bands at K and K’ (see Fig. \ref{fig1}) \cite{xiao_coupled_2012, kormanyos_monolayer_2013, Bawden2016}. This peculiar spin-orbit interaction has led to new developments in the field of non-conventional superconductivity, as recent measurements on monolayer and few layers NbSe$_2$ samples showed that the in plane critical magnetic field is much larger than the paramagnetic limit\cite{Grigorieva, xi_ising_2016}. 
It has been shown that both  phase transitions can be partially tuned by applying a gate
voltage \cite{xi_gate_2016, lu_evidence_2015}. It is thus crucial to study the evolution of charge ordering and superconductivity in 2D metallic dichalcogenides in the limit of 
heavy electron-doping, going way beyond what can be achieved by FET doping techniques (see Fig. \ref{fig1}).

Most of the studies on few layers NbSe$_2$  were done using mechanical\cite{Novoselov10451} or liquid exfoliation \cite{Coleman568} which implies finding a flake of sample of the appropriate thickness among the many pieces dispersed on a substrate. Doping is then achieved via FET or adatom deposition. Here, we propose an alternative technique that allowed us to investigate the extreme high-doping limit of a NbSe$_2$ single layer. We propose to use a compound belonging to the misfit TMDs. These misfits materials consist of transition metal dichalcogenides layers sandwiched with rock salt chalcogenides layers \cite{rouxel_chalcogenide_1995}, namely (RQ)$_{1+x}$(TQ$_2$)$_n$ (R=Pb, Bi or Rare-Earth, Q=Chalcogene, T=Transition metal and $n = 1, 2$) (see Figs. \ref{fig1} and \ref{fig2}) which exhibit an intergrowth structure formed by the alternated stacking of  RQ planes (having NaCl structure) with TMD layers (CdI$_2$ or NbS$_2$ structure).  An electronic  charge transfer occurs from the RQ to the TMD layers stabilising the structure. It was suggested that these compounds could be considered as intercalated compounds, however it is unclear if  a rigid band model of doping could apply \cite{cario_stability_1997, cario_band_1999-1}.  It is possible to obtain a large variety of multilayers of TMDs (1, 2 or 3 layers) separated by insulating RQ layers. For instance, the (LaSe)$_{1.14}$(NbSe$_2$)$_2$ compound is composed by metallic 2H-NbSe$_2$ bilayers separated by insulating LaSe bilayers (Figs.\ref{fig1} and \ref{fig2}).  It is even possible to use multilayers of RQ in order to get quasi-2D non coupled NbSe$_2$ multilayers, thus realizing nearly ideal embedded 2D systems with a well-defined multilayer structure.

Many misfits layered compounds were found to be superconducting, like (PbS)$_{1.14}$NbS$_2$ or (LaSe)$_{1.14}$(NbSe$_2$)$_2$  \cite{rouxel_chalcogenide_1995} with $T_c = 2.5$~K and $T_c =5.3$~K  respectively. (LaSe)$_{1.14}$(NbSe$_2$)$_2$ displays an in-plane critical field around 30~T way beyond the 11.5~T  paramagnetic limit required to break singlet Cooper pairs. A similar compound (LaSe)$_{1.14}$(NbSe$_2$), consisting of a stacking of 1H-NbSe$_2$ and LaSe single layers, has a critical temperature of 1.2~K and a huge in-plane critical field of more than 16~T, which is well beyond the paramagnetic limit of 2.6~T \cite{monceau_anisotropy_1994, kacmarcik_upper_2000, kacmarcik_scanning_2004, sofranko_periodic_2020}. This enhanced critical field is similar to what was reported for few layers NbSe$_2$ \cite{xi_ising_2016} and gated MoS$_2$ \cite{lu_evidence_2015}.  These unusual superconducting properties indirectly suggest that the material behaves similarly to monolayer NbSe$_2$ making it a bulk yet quasi-2D system. However, up to now there is no direct evidence of its bidimensionality and no measurement of the amount of doping in NbSe$_2$ layers. 

Here we show  that (LaSe)$_{1,14}$(NbSe$_2$)$_2$ hosts indeed a band structure that is intimately related to the one of a strongly and rigidly doped NbSe$_2$ monolayer. We measured finely the Fermi surface using quasiparticles interferences (QPIs) spectroscopy and ARPES. Our measurements, supported by extensive first-principles DFT calculations, demonstrate that (LaSe)$_{1,14}$(NbSe$_2$)$_2$ behaves as a heavily electron-doped monolayer NbSe$_2$ with a Fermi level shifted of +0.3~eV, or equivalently $\approx 0.6$ electron per niobium atom. 

The general idea of our work is summarized in Fig. \ref{fig1} (b) where the band structure of 1H-NbSe$_2$ monolayer is shown in black and the chemical potential of the undoped monolayer  in blue. The gray ribbon labels the range of doping achievable with ionic liquid gating\cite{MakGatedNbSe2}. It has to be compared with the level of doping achieved in the misfit compound via electronic charge transfer from the LaSe to the NbSe$_2$ layers, shown in yellow. This tremendous doping is several times larger than what can be achieved by conventional ionic liquid doping in field effect transistor geometry. As a consequence, the shape of the Fermi surface is greatly modified and in particular the Fermi pockets at $\Gamma$ and K are strongly shrunk as our quasiparticle interferences and ARPES measurements reveal. The ability to induce a huge doping such as to change dramatically the shape of the Fermi surface is extremely promising for future work on electronic properties of misfit compounds. 

\section{Structure}

\begin{figure*}[h!]
\begin{center}
\includegraphics[width = 14 cm]{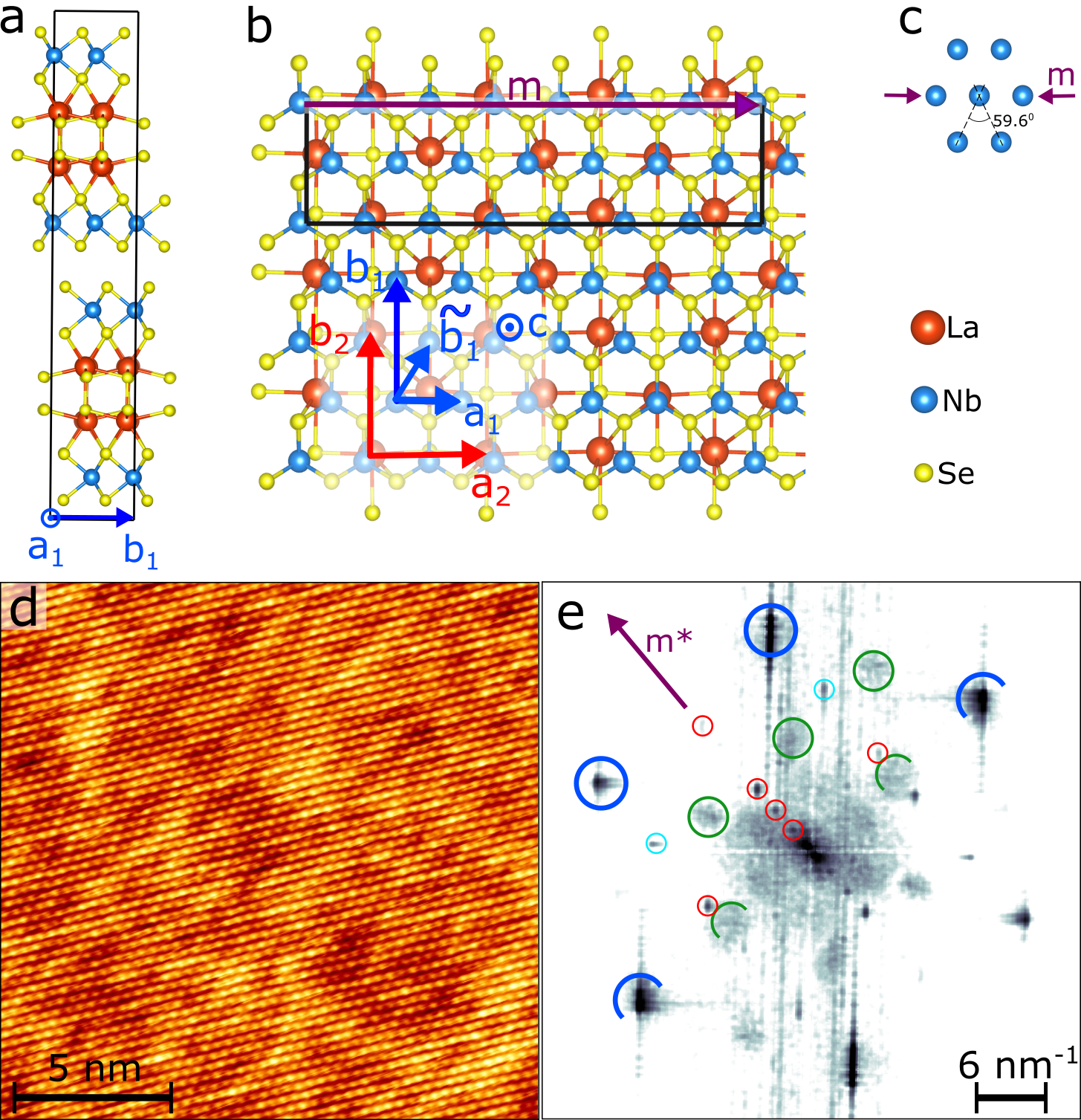}
\caption{In (a) stacking along c axis of (LaSe)$_{1.14}$(NbSe$_2$)$_2$. The material can be seen like a van der Waals stacking of units composed of a LaSe layer sandwiched by two NbSe$_2$ monolayers. Each NbSe$_2$ layer is strongly bound to LaSe by iono-covalent bonding. A top view is shown in (b), the black rectangle represents a nearly commensurate approximate of the crystal structure with $m\approx 7 a_1\approx 4 a_2$. The lattice of the NbSe$_2$ layers is not perfectly hexagonal, it is slightly compressed along the $\vec{b}_1$ direction as shown in (c).In (d), STM topography of a cleaved  (LaSe)$_{1.14}$(NbSe$_2$)$_2$ sample measured at 4.2~K (V$_T=-200$~mV). The triangular atomic lattice indicates that the sample is terminated by a NbSe$_2$ layer. In (e) Fourier transform modulus of the topography. The dark blue circles indicate the NbSe$_2$ Bragg peaks and the red circles highlight some incommensurability peaks due to the modulation of the NbSe$_2$ lattice by the underlying LaSe lattice.  The light blue circles correspond to  the underlying LaSe lattice. The green circles are associated to a $2 \times 2$ charge modulation with a short coherence length.  }

\label{fig2}
\end{center}
\end{figure*}

The misfit material (LaSe)$_{1,14}$(NbSe$_2$)$_2$ is a layered compound consisting of a regular alternation of 2H-NbSe$_2$ bilayers with trigonal prismatic structure and of rocksalt two atom thick LaSe layers (Figs. \ref{fig1}-\ref{fig2}). Each NbSe$_2$ and LaSe sublattice has its own set of cell parameters noted $a_v$, $b_v$ and $c_v$ with \(v\)=1 and 2, respectively. Compared to bulk 2H-NbSe$_2$ the lattice of NbSe$_2$ layers is not perfectly hexagonal and is slightly compressed along the $\vec{a}_1$ direction (space group C$_{2221}$, see Fig. \ref{fig2}(c)). As a consequence the NbSe$_2$ sublattice is described by a centered orthorhombic cell with in plane lattice vectors $a_1=3.437$~\AA\ and $b_1 \approx 6$~\AA. The LaSe sublattice has also an orthorhombic symmetry but with similar in plane lattice parameters $a_2 \approx b_2\approx 6$~\AA. Both NbSe$_2$ and LaSe layers have the same $\vec{b}$ lattice vector ($\vec{b}_1$=$\vec{b}_2$) and the material is commensurate in this direction. However, the ratio between the norms of the $\vec{a}_1$ and $\vec{a}_2$ vectors, sharing a common direction, is an irrational number ($a_2/a_1=1.751\approx 7/4$), making (LaSe)$_{1,14}$(NbSe$_2$)$_2$  incommensurate in the $\vec{a}$ direction. However, $a_2/a_1 \approx 7/4$ such that (LaSe)$_{1,14}$(NbSe$_2$)$_2$  can be considered as almost periodic in the $\vec{a}_2$ direction with an approximate commensurate lattice vector $\vec{m}$ ($m = 7 a_1 \approx 4 a_2$).

 A commensurate structure with $m = 7 a_1 \approx 4 a_2$ was assumed in first principles electronic structure calculations (see the method section). We performed structural optimization for the full bulk structure (232 atoms per cell). We used different starting guesses obtained by superimposing in different ways LaSe and NbSe$_2$  bilayers (see SI). Minimization of the total energy gives two inequivalent minima that however result in very similar
structures with practically identical electronic structures.
Very strong iono-covalent bonds form at the contact between NbSe$_2$ and LaSe with a substantial deformation of the LaSe layer (see Fig. \ref{fig2} and
movie in SI) and a significant deviation from an ideal rock salt structure. This huge modification of the LaSe interface signals a large
charge transfer between LaSe and NbSe$_2$ bilayers.
On the contrary the van der Waals interface between the 
two NbSe$_2$ layers is weakly affected.
This explains why the cleavage of the sample leads to a 
monolayer (ML) NbSe$_2$ terminated surface (see Figs. \ref{fig1}-\ref{fig2}) that we probed by STM and ARPES.

A typical STM topography map of (LaSe)$_{1,14}$(NbSe$_2$)$_2$ surface of an in situ cleaved sample is represented in Fig. \ref{fig2} (d). Since the surface layer is 1H-NbSe$_2$ , the atomically resolved STM images show the Se atoms of NbSe$_2$. The rectangular centered Se lattice appears almost hexagonal and for simplicity, in the following, the NbSe$_2$ layers will be referred as hexagonal. In addition to the Se atomic lattice, one observes multiple additional spatial modulations which are due to the incommensurability with the underlying LaSe layer.  In order to properly characterize the origin of these modulations, the modulus of the Fourier transform of a similar but larger scale topographic image taken on the same area with same scanning angle is represented in Fig. \ref{fig2} (e). The Bragg peaks of the reciprocal lattice of NbSe$_2$ are highlighted by blue circles. The sharp peaks  circled in red are the first, second, third and sixth harmonics of the approximate periodic misfit potential. Additional peaks attributed to the LaSe lattice are also observed and accented by light blue circles. The signals presented so far can directly be deduced from the theoretical crystalline structure. In addition to those, one can also observe other contributions. First, diffuse peaks (highlighted by green circles) lie at half the Bragg peaks of the NbSe$_2$ hexagonal lattice. They express the presence a $2\times 2$ charge modulation in the system. This charge order will be fully discussed later in this paper. In addition, a halo surrounding $\Gamma$ quasi-reaching half the Bragg peaks can be seen. It is attributed to integrated quasiparticle interference spectroscopic signal as we discuss below.

\section{Electronic band structure}

\begin{figure*}[h]
\begin{center}
\includegraphics[width = 14 cm]{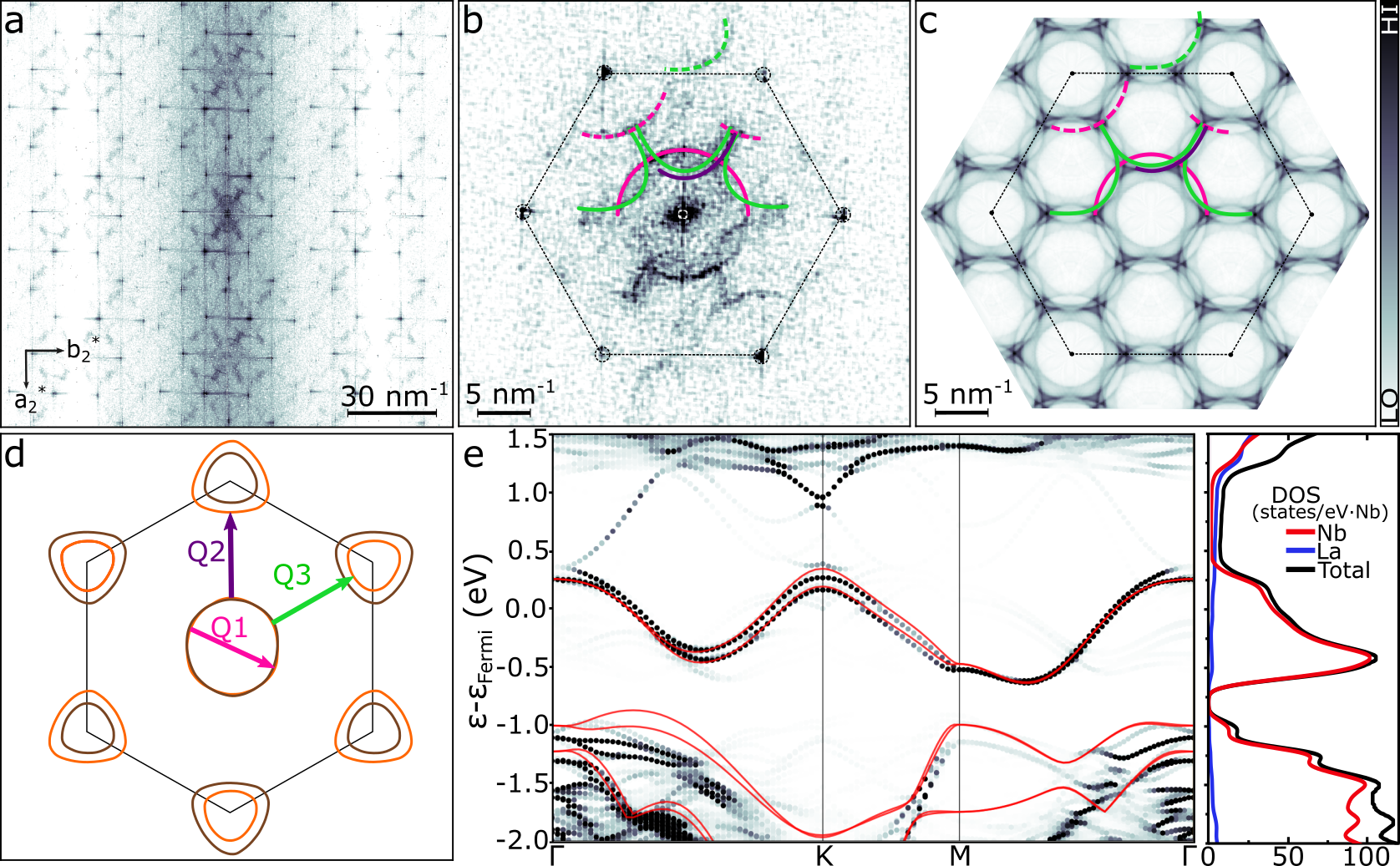}
\caption{In (a) Fourier transform modulus of a highly resolved STM topography of (LaSe)$_{1,14}$(NbSe$_2$)$_2$ ($T= 4,2$~K, $V_T=-20$~mV) showing QPIs over several NbSe$_2$ Brillouin zones. The signal in dark gray is stronger in the $b_1^*$ direction which corresponds to the direction of the incommensurability. In (b) Fourier transform of a conductance map at Fermi energy ($T= 4,2$~K). A DFT simulation of the QPIs is shown in (c), with the same highlighted details as in (b). (d) Fermi Surface of a 1H-NbSe$_2$ monolayer for the doping level found in  (LaSe)$_{1,14}$(NbSe$_2$)$_2$. The different color of the Fermi surface at $K$ and $K'$ corresponds to the different out-of-plane spin polarization. A few scattering channels implied in the QPI images are shown with arrows. In (e) the unfolded electronic band structure of (LaSe)$_{1,14}$(NbSe$_2$)$_2$ is shown in gray dots, the dot color scale following the spectral weight of the band. It is compared with the dispersion of a NbSe$_2$ monolayer (red curve). The right panel of (e) displays the contribution from La and Nb atoms to the density of states.    }

\label{fig4}
\end{center}
\end{figure*}

We then study the electronic structure of the surface of  NbSe$_2$ terminated (LaSe)$_{1,14}$(NbSe$_2$)$_2$ 
by using DFT calculations. The results are presented in Fig. \ref{fig4} (e) after band unfolding\cite{bandsup1,bandsup2} onto the NbSe$_2$ three-atom C$_{2221}$ pseudo-hexagonal unit-cell having $a_1=3.457$ \AA\ and $\tilde{b}_1= 3.437$ \AA  ($\tilde{b}_1=(a_1+b_1)/2$, see Fig.\ref{fig2} (b)).  As a reference, we also plot in red the electronic structure of an isolated  1H-NbSe$_2$ layer with the same $a_1$ and $\tilde{b}_1$ but doped by $0.6$ electrons per Nb atom (we use a compensating positive background to ensure charge neutrality). As it can be seen, the electronic structure of (LaSe)$_{1,14}$(NbSe$_2$)$_2$ in a $\pm 0.5$ eV range from the Fermi level is almost indistinguishable from that of a rigidly doped NbSe$_2$ monolayer. This is further confirmed by the detailed comparison of the electronic structure of the (LaSe)$_{1,14}$(NbSe$_2$)$_2$ surface with that of a rigidly-doped single layer NbSe$_2$ with C222$_1$ unit cell (see Figs. S1 and S2 in SI). The only differences
occur at zone center where a lanthanum band hybridizes with the single
layer NbSe$_2$ band. The resulting band deformation leads to a  small
La component at the Fermi level ($\epsilon_F$) close to  zone center. The amount of La electrons at $\epsilon_F$  remains, however, negligible, 
as demonstrated by  the density of states projected over La and Nb 
atomic states in the rightmost panel of Fig. \ref{fig4} (e).
Finally, we underline that, even if we keep in all simulations the  non-perfect hexagonal symmetry due to the 
anisotropy in $a_1$ and $\tilde{b}_1$, this effect is completely
irrelevant and does not lead to any sizeable effect in the electronic structure and Fermi surface. The calculated Fermi surface of the isolated layer is
plotted in Fig. \ref{fig4} (d) with the degree of spin up and down polarization illustrated in different colors.

We probed the band structure at Fermi energy by
quasiparticle interference spectroscopy measurements on 
NbSe$_2$ terminated (LaSe)$_{1,14}$(NbSe$_2$)$_2$ .
This technique relies on the scattering of the electron Bloch functions by defects. The incoming and scattered electrons may give rise to some interference pattern that can be measured by scanning tunneling spectroscopy.
In Fig.~\ref{fig2}(e) some QPIs are already appearing, but they are not well resolved. This is due to the fact that the topography was recorded with a bias of 200~mV and, as such, it is sensitive to the integral of the local density of states in an energy window of 200~meV. Thus the spectroscopic halo around $\Gamma$ in Fig.~\ref{fig2}(e) is actually a mix between the QPIs patterns at all energies ranging from 0~meV to -200~meV. In order to get a highly resolved QPIs measurement of the band structure at the Fermi energy we performed topography measurements with a low voltage of 20~mV so that only electrons  in a 20~meV window close to $\epsilon_F$ contributed (Fig. \ref{fig4} (a)). As  the band dispersion over a 20~meV range is negligible (see Fig. \ref{fig4} (e)),  these measurements should reflect the joined density of states at the Fermi level (see method section). The Fig. \ref{fig4} (a) shows a Fourier transform of the topography which spans many Brillouin zones. One can observe a dark vertical band which indicates that a stronger QPI signal is generated in the direction of the incommensurability.  This suggests that the incommensurability plays the role of electron scattering potential. Moreover, as atomically resolved images in Fig. \ref{fig2} do not present any visible defects such as vacancies, steps or dislocations (i.e. the samples are very clean), the only scattering mechanism susceptible to generate QPIs is provided by the incommensurability.

In order to get more insight on the band structure we performed QPIs measurements by taking the squared modulus of Fourier transform of conductance maps measured at the Fermi energy. Such a QPIs measurement is shown in Fig. \ref{fig4} (b), where one can see a complex structure with some almost circular patterns. The different circular patterns are associated to different scattering mechanisms shown in Fig.  \ref{fig4} (d). For instance the $Q_1$  vector depicts scattering events from the $\Gamma$ pocket to itself, while $Q_2$ depicts scattering between the $\Gamma$ pocket and the outer K or K$^{\prime}$ pockets. Finally, the vector $Q_3$ schematizes scattering in between $\Gamma$ and the internal K and K’ pockets. These different scattering channels lead to the nearly circular patterns depicted in color on Fig. \ref{fig4} (b). 
We calculated the QPI patterns (see methods and SI) by using DFT
for a three atoms NbSe$_2$ unit cell for an unsupported monolayer with a doping of $\approx 0.55-0.6$ electrons per Nb atom. 
The simulated QPI, shown in Fig. \ref{fig4} (c) exhibits the same circular patterns as the experiment. In order to compare quantitatively the experimental and theoretical QPIs we plotted an identical set of circles on both Figs. \ref{fig4} (b) and (c), where the color corresponds to different scattering channels as explained above. The agreement is very good; all the patterns observed experimentally are accurately reproduced by the rigid band approximation of  1H-NbSe$_2$ monolayer doped by $\approx 0.55-0.6$ electrons per Nb, except some spectral weight at the Bragg peaks. This latter electron doping agrees well with the chemical estimate that, in (LaSe)$_{1,14}$(NbSe$_2$)$_2$, the (LaSe)$_{1,14}$ should provide 1.14 electrons to (NbSe$_2$)$_2$. This simple chemical intuition leads to 0.57 electron doping which is in very good agreement with DFT and experiments.

One of the major interests of single layers 1H TMDs is their Ising spin orbit splitting.  This spin orbit splitting should manifest with two scattering channels corresponding to the scattering from $\Gamma$ to the outer K pocket and from $\Gamma$ to the inner K pocket, represented schematically by $Q_2$ and $Q_3$, respectively in Fig. \ref{fig4} (d). This should lead to two concentric circles schematized by the purple and green arcs in Figs.\ref{fig4} (b) and (c). This splitting is indeed observed even though the signal to noise ratio is pretty weak. 
 
\begin{figure*}[h]
\begin{center}
\includegraphics[width = 14 cm]{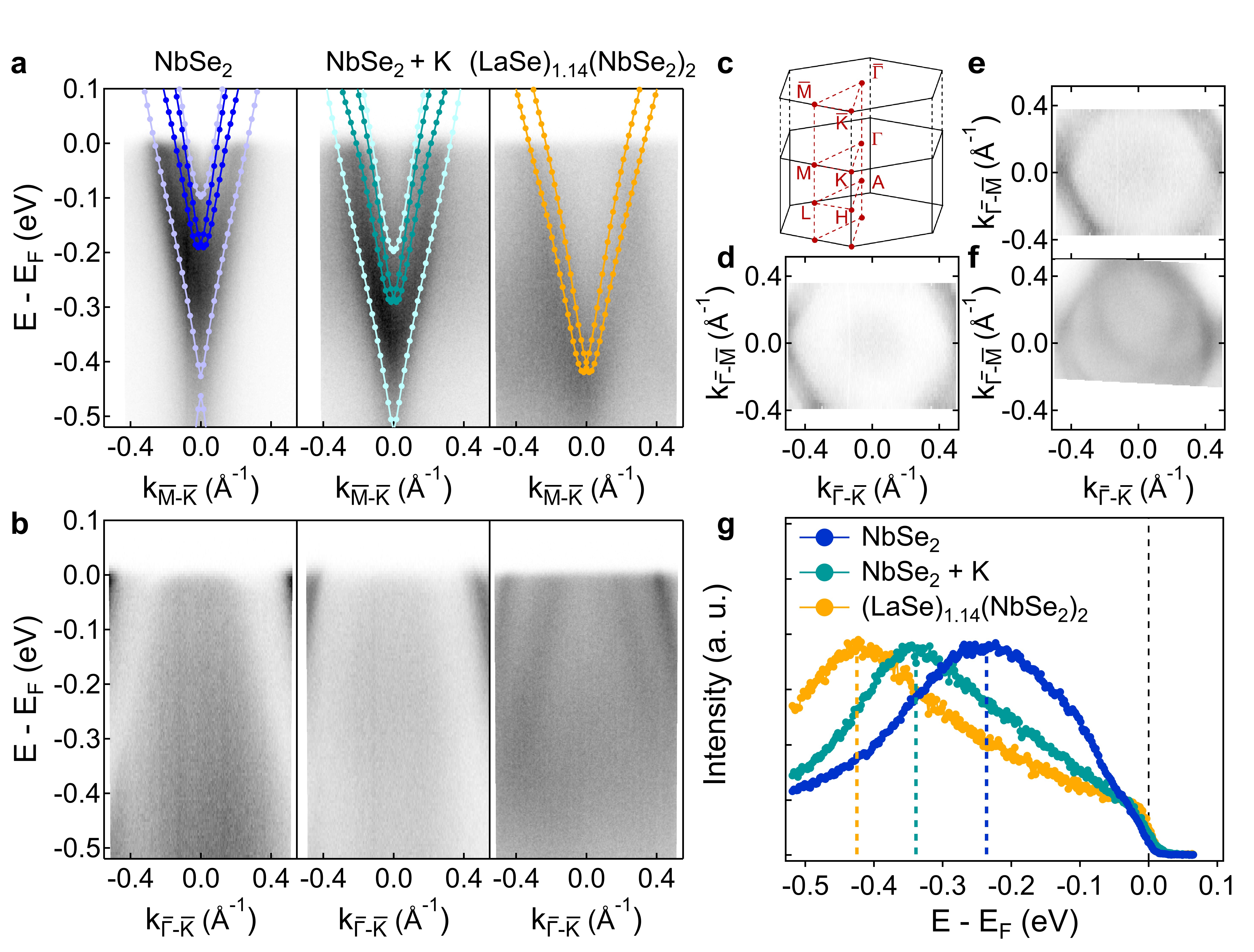}
\caption{a) From left to right: ARPES spectra along the $\bar{K'}$-$\bar{M}$-$\bar{K}$ high-symmetry line of the Brillouin zone of pristine NbSe$_2$ (left), K-doped NbSe$_2$ (middle) and of the misfit compound (LaSe)$_{1.14}$(NbSe$_2$)$_2$, measured using HeI photon energy ($h\nu$=$21.2$ eV) and at T = 50 K. On the left panel the pristine NbSe$_2$ band structure calculated by DFT is show in light blue for $k_z=0$ and dark blue for $k_z=1/2$. In the middle panel the same DFT curves are shown shifted from 0.1~eV. In the right panel, the orange doted curve represents the DFT calculated band structure of a monolayer NbSe$2$ with a doping of $0.55-0.6$ electrons per Nb atom. b) Same as a) along the $\bar{K'}$-$\bar{\Gamma}$-$\bar{K}$ high-symmetry line of the Brillouin zone. c) Surface and three-dimensional Brillouin zone of 2$H$-NbSe$_2$. d)-f) Fermi surfaces around the $\bar{\Gamma}$ point of the Brillouin zone of pristine NbSe$_2$ (d), K-doped NbSe$_2$ (e), and of the misfit compound (LaSe)$_{1.14}$(NbSe$_2$)$_2$ (f). g) Energy distribution curves taken at $\bar{M}$ on a) for pristine NbSe$_2$ (blue), K-doped NbSe$_2$ (green) and of the misfit compound (LaSe)$_{1.14}$(NbSe$_2$)$_2$ (orange). The coloured vertical dashed lines indicate the energy position of the maxima of the EDCs corresponding to the bottom of the electron pockets at $\bar{M}$ associated with the three different samples.} 

\label{fig5}
\end{center}
\end{figure*}

According to our QPIs measurements and DFT calculations we found that the band structure of the NbSe$_2$ terminated surface of (LaSe)$_{1,14}$(NbSe$_2$)$_2$  is well accounted for by a rigid band approximation with a huge Fermi level shift. In order to ascertain this finding, we carried out some comparative ARPES measurements of the low-energy electronic states of bulk 2H-NbSe$_2$, K-doped 2H-NbSe$_2$ (see Methods) and (LaSe)$_{1.14}$(NbSe$_2$)$_2$ at T=50 K and using He-I radiation ($h\nu$=$21.2$ eV). Figures \ref{fig5} (a) and (b)  show the corresponding ARPES intensity maps along the $\bar{K'}$-$\bar{M}$-$\bar{K}$ and $\bar{K'}$-$\bar{\Gamma}$-$\bar{K}$ high-symmetry lines of the hexagonal surface Brillouin zone of 2H-NbSe$_2$ (Fig. \ref{fig5} (c)), respectively. Around $\bar{M}$, for pristine 2H-NbSe$_2$ (Fig. \ref{fig5}, left panel), we can easily recognize the spin-polarized Nb 4$d$-derived states coming from the cut of the two strongly trigonally-warped barrels of the Fermi surface centred around each zone-corner $\bar{K}$ ($\bar{K'}$) point (also see the calculated  Fermi surfaces in Fig. \ref{fig4}) \cite{Bawden2016}. 
At $\bar{\Gamma}$ (Fig. \ref{fig5} (b), left panel), whereas the Nb-dominated states yield strong spectral weight close to the Fermi level crossings away from normal emission, additional diffuse spectral weight coming from a highly three-dimensional Fermi surface sheet of predominantly Se $p_z$ orbital character is visible at the zone centre due to the finite out-of-plane momentum resolution of ARPES (also see the measured Fermi surface Fig. \ref{fig5} (d)) \cite{Bawden2016}.

The same spectral features are also seen for the K-saturated surface of NbSe$_2$ with a yet higher diffuse background (coming from the randomly adsorbed alkali atoms at the NbSe$_2$ surface) and a different chemical potential as evidenced by the shift towards higher binding energies of both the electron pocket at $\bar{M}$ and the hole pocket at $\bar{\Gamma}$ (Fig. \ref{fig5} (a) and (b), middle panels) and the reduced area of the $\bar{\Gamma}$-centred Fermi surface (Fig. \ref{fig5} (e)). As at the photon energy used here, the inelastic mean free path of the photoelectrons is on the order of the interlayer distances, we are predominantly sensitive to the doping of the topmost layer of the unit cell. From our ARPES measurements, we can thus conclude that the K-saturated NbSe$_2$ surface has been electron-doped within a rigid-band picture corresponding to a chemical potential shift of $\approx$ 100 meV. 

Focusing now on (LaSe)$_{1.14}$(NbSe$_2$)$_2$, we observe further energy shifts towards higher binding energy of both the electron and hole Nb-derived pockets (Fig. \ref{fig5} (a) and (b), right panels) as well as a still reduced Fermi surface area (Fig. \ref{fig5} (f)). This indicates that the electron-doping of the NbSe$_2$ layer at the interface with LaSe within the misfit compound is much higher than the one obtained by alkali gating. Energy distribution curves taken at the $\bar{M}$ point for bulk NbSe$_2$, K-doped NbSe$_2$ and (LaSe)$_{1.14}$(NbSe$_2$)$_2$ (Fig. \ref{fig5} (g)) demonstrate that such a doping is almost twice the one reached in K-doped NbSe$_2$ with a chemical potential shift of $\approx$190 meV (see the vertical dashed lines on Fig. \ref{fig5} (g) showing the energy position of the bottom of the electron pockets at $\bar{M}$) with respect to pristine NbSe$_2$.  Interestingly, we can also note the absence of spectral intensity coming from the Se $p_z$ orbital at the zone centre (Fig. \ref{fig5} (b). right panel), indicating that the band structure of NbSe$_2$ within the misfit compound is close to the one of a strongly electron-doped NbSe$_2$ \textit{monolayer}. Nevertheless, the electrons feel a rather strong competing potential coming from the intrinsic incommensurability of the heterostructure as evidenced by the clear presence of shadow bands not only in the ARPES intensity maps but also at the Fermi surface (Fig. \ref{fig5} (f)).

\section{emergence of $2\times 2$ charge ordering}

Bulk 2H-NbSe$_2$ shows a charge density wave (CDW) transition at around 35~K with a nearly commensurate $3\times 3$ charge order\cite{WilsonDisalvo}. In single layer, contradictory results have been reported as 
ML NbSe$_2$ on top of Graphene shows a strong increase
of T$_{CDW}$ with respect to the bulk\cite{xi_strongly_2015}, while ML NbSe$_2$ grown on top of SiO$_2$ shows a $3\times 3$ CDW with a similar T$_{CDW}$ as in the bulk. Recent first principles calculations\cite{BiancoNbSe2} confirm the second behaviour. It has been shown that electron doping
by FET strengthen T$_{CDW}$ \cite{xi_gate_2016}, it is however unclear if a change of ordering vector occurs. The continous T$_{CDW}$ curve versus FET doping in Ref\cite{xi_gate_2016} seems to suggest that it is not the case, however the range of considered doping is much lower. 

\begin{figure*}[h]
\begin{center}
\includegraphics[width = 16 cm]{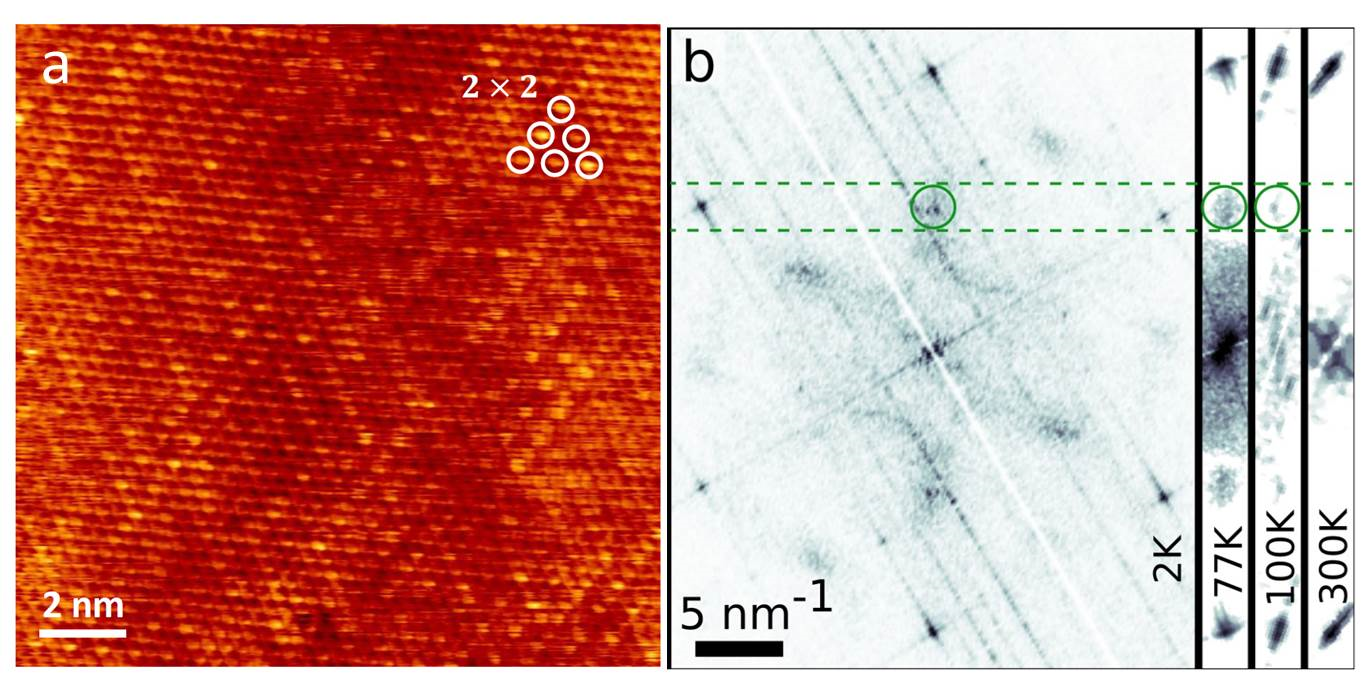}
\caption{In (a) STM topography evidencing $2\times 2$ charge modulation with a short coherence length ($T=2$~K, $V_T=24$~mV). The corresponding Fourier transform modulus is shown in (b). The right panel in (b) shows the evolution of the Fourier transform modulus with temperature. The $2\times2$ pattern is still visible at 100~K, it disappears at $\approx105$~K. } 

\label{fig6}
\end{center}
\end{figure*}

Since (LaSe)$_{1,14}$(NbSe$_2$)$_2$ shows an energy band structure analogous to 1H-NbSe$_2$ monolayer one expects CDW to show up at low temperatures. It is well established that CDW are in general very sensitive to the shape of the Fermi surface. Thus, one expects that the strong doping of $0.55-0.6$ electrons/Nb could induced some change in the CDW periodicity.  This is indeed the case as we found no experimental evidence of a $3\times 3$ charge ordering, but we find $2\times2$ patches appearing  below approximately 105~K in most parts of the sample. 
Our experimental observations are summarized on Fig. \ref{fig6}, where a topography with a $2\times 2$ charge order is shown in panel (a), while the Fourier transform modulus of the topographies measured at various temperature are shown in (b). The Fourier transform shows a six-fold peak structure at half-Bragg positions that corresponds to a $2\times 2$ charge modulation in real space. The $2\times 2$ signal is still present in Fourier map at 100~K and it disappears above 105~K see Fig. \ref{fig6} (b).  However, the $2\times2$ reconstruction does not cover
completely the sample in a uniform way, but is more
clearly visible in some regions with respect to others.
The corresponding CDW peak in Fig. \ref{fig6} is also less sharp, probably meaning that regions of $2\times2$ ordering coexist with patches with no CDW ordering. 

In order to investigate the possible evolution of T$_{CDW}$ with doping, we performed harmonic phonon calculations using density functional perturbation theory
\cite{QE1,QE2} for a single layer NbSe$_2$ as a function of electron doping. The system is kept neutral by using a positive jellium background.
The hexagonal-compression of the lattice is included, so that there are two inequivalent $M$ points (labeled $M$ and $M^{\prime}$). The results are shown in SI, Fig. S3. At zero doping, we find a broad region of instability close to $2/3 \Gamma M$ and $2/3 \Gamma M^{\prime}$, compatible with a $3\times 3$ CDW, signaled by the occurrence of an imaginary mode both along $\Gamma M$ and $\Gamma M^{\prime}$. At a doping of $0.4$ electrons/Nb, the region of instability shifts at the $M$ and $M^{\prime}$ points, in qualitative agreement with experimental findings. It is known that \cite{BiancoNbSe2} the CDW in NbSe$_2$ arises from
an interplay between Fermi surface effects and ionic fluctuations related to anharmonicity. Fermi surface effects determine the ordering vector, while the occurrence of the CDW is mostly related to phonon-phonon scattering. In our calculation we only include harmonic terms, thus we overestimate the tendency towards CDW. At a doping of $0.4$ electrons/Nb, the ordering vector changes
essentially because the Fermi surface geometry is affected and a nesting compatible to $2\times 2$ reconstruction occurs. 
However, at the doping of $0.55-0.6$ electrons/Nb, that is the one estimated by comparing our electronic
structure with QPI and ARPES, no instability is found, mainly because 
the vector $q=0.5\Gamma_M $ does not connects efficiently portion of the Fermi surface. Anharmonicity would suppress even more the CDW instability.
A possible way to reconcile theory and experiments, 
is to consider a non-uniform doping in the sample and the possible existence of patches with lower doping. These
regions are not detected in ARPES and QPI that probe larger regions of the sample, however they can be detected by the ultrafine spatial sensitivity of STM.

\section{Conclusion}
The success of two dimensional crystals in the fields of physics, nanotechnology, chemistry and material science is largely due to 
the much greater tunability of their properties with respect to their bulk counterparts. The three main parameters that can be varied are the sample thickness, 
the Moir\'e angle between the stacked layers and doping.
However, while the first two 
parameters can be varied at will, for NbSe$_2$ doping is limited to $\approx 10^{14}$ electrons cm$^{-2}$\cite{MakGatedNbSe2}, that is the maximum amount of carriers that can be introduced in an electrical double-layer field-effect
transistor geometry. Somewhat larger doping can be achieved via deposition of Alkali atoms (such as potassium). However, in many cases, the deposition of alkali atoms on a 2D crystal is difficult to achieve as the coverage can be non-uniform or the wetting of the 2D crystal by the adatoms not even possible. 

Here we propose to use rocksalt rare-earths chalcogenides sandwiched with transition metal dichalcogenides to build a misfit compound. Both the thickness of the transition metal dichalcogenide and of the rocksalt rare-earth chalcogenide can be tuned at will \cite{wiegers_misfit_1992, wiegers_misfit_1996-1},
exactly as in a 2D crystal. The large electron charge transfer from the rare-earth chalcogenide to the TMD  allows to obtain doping fractions that largely encompass both those obtained in the case of ionic-liquid-based field-effect transistors and K adatom deposition. We have shown that this is indeed the case for NbSe$_2$ bilayers sandwiched in rocksalt LaSe for which we  achieve a doping of $0.55-0.6$ electrons per Nb, corresponding to a $\approx 0.3$ eV shift of the Fermi level.
This large electron  transfer from LaSe to the TMD is due to the chemical properties of the highest occupied states of  bulk LaSe that are formed only by atomic La states\cite{osti_1188105}. As a consequence LaSe essentially behaves as a donor, an effect that we expect not to depend on the kind of TMD used to build the misfit structure. This suggests 
 that similarly large doping can be achieved by sandwiching LaSe with  other metallic TMDs.

Finally, we remark that it is also possible to vary the rocksalt layer, opening a tremendous space of possibilities and a fertile ground for the discovery of highly innovative materials. For instance, assuming La$^{3+}$, Pb$^{2+}$ and Se$^{2-}$ in the hypothetical series of compounds (La$_{1-y}$Pb$_y$Se)$_{1+x}$(NbSe$_2$)$_2$ a simple charge balance calculation shows that the charge transfer from the rock salt layer (La$_{1-y}$Pb$_y$Se) to the (NbSe$_2$)$_2$ double layer could vary continuously from $0$ to $1+x$ by reducing the lead content. While the (La$_{1-y}$Pb$_y$Se)$_{1+x}$(NbSe$_2$)$_2$ series was never reported so far, similar substitutions of La$^{3+}$ per Sr$^{2+}$ substitution were already demonstrated in the (LaS)$_{1+x}$VS$_2$ \cite{nishikawa_compounds_nodate} and (LaS)$_{1+x}$CrS$_2$ \cite{cario_stability_1997}. Therefore, the charge transfer could be tuned by appropriate substitution in misfit layer compounds, pretty much in the same way as doping can be varied in field-effect transistor, but in a much broader range.

We have shown that the effect of a large electron doping in single layer NbSe$_2$
results in the change of the ordering vector of the
charge density wave. However, other properties can also be explored in 
misfit compounds, such as  topological and Ising superconductivities in NbSe$_2$ and MoS$_2$,
 Mottronics with misfit compounds involving  1T-TaSe$_2$ or 1T-NbSe$_2$ layers \cite{janod_resistive_2015, nakata_monolayer_2016, colonna_mott_2005} or thermoelectric applications \cite{xu_enhanced_2014, rameshti_thermoelectric_2016, qian_quantum_2014} where the ability to fine tune the chemical potential in stacks of 2D topological insulators such as 1T$^{\prime}$-WSe$_2$ would be a perfect tool for rational design of materials with high figures of merit.

\section{Methods}

Single crystals of (LaSe)$_1.14$(NbSe$_2$)$_2$ were prepared by the solid-state reaction of the elemental precursors (i.e. La, Nb, Se) and subsequent chemical vapor transport using I$_2$. Under inert atmosphere, millimeter-sized La powder was freshly scraped from the ingot (Strem Chemicals, 99.9\%) and mixed with Se (Alfa Aesar, 99.999\%) and Nb powder (Puratronic, 99.99\%) in the molar ratio of La / Nb / Se = 1.14 / 2 / 5.40. Here Se was added in 5\% excess over the nominal stoichiometry La / Nb / Se = 1.14 / 2 / 5.14. The mixture was manually ground in an agate mortar and transferred in a silica tube (~15 cm), which was subsequently evacuated to 10$^{-3}$ torr and sealed by flame. The sealed mixture was firstly heated to $200^\circ$C at a rate of $50^\circ$C~h$^{-1}$ and held for 12~h. Then the temperature was raised to $900^\circ$C at the same rate and heated for 240~h before the furnace was subject to radiative cooling. The lustrous black powder (ca. 500 mg) obtained from the reaction was placed in the clean silica tube (length: 15cm) together with 43 mg of iodine (Aldrich, 99.9\%), followed by the evacuation at liquid nitrogen temperature (77~K) and flame sealing. The reaction mixture was loaded into the furnace designed to create temperature gradient: the lump of the mixture was heated at $900^\circ$C on one side of the tube and the other side was held around $750^\circ$C. After 240~h of the thermal treatment, the reaction was quenched by plunging the tube into water. The single crystals grown on the wall of tube were washed with water and ethanol and stored in vacuum. The compositional integrity of the obtained crystals was checked by their backscattered electron images and energy-dispersive X-ray (EDX) spectra acquired on scanning electron microscopy (JEOL JSM 5800LV).

In order to access the LDOS at different energies, spectroscopic $I(V)$ maps were acquired, consisting in taking individual $I(V)$ spectra at many positions determined by a predefined spatial grid. The differential conductance $dI/dV$  maps (derived afterwards) at a given voltage V can directly be related to the LDOS of the sample at the corresponding energy. Performing LDOS experiments at 4.2~K allowed us to exhibit QPIs patterns which give access to the Fermi surface of (LaSe)$_{1,14}$(NbSe$_2$)$_2$.
The study was realized with an Omicron LT STM and an home-made 300~mK STM using Pt80\%/Ir20\% tips. Prior to each experiment, the DOS of the tip was verified on Pt and HOPG samples and checked to be featureless. The topography maps were taken in constant current mode and the spectra were taken at constant tip-sample distance for a given tunneling set point (mainly around $U=-20$~mV, $I=100$~pA). The samples were cleaved at room temperature in ultrahigh vacuum under a pressure of approximately 5.10$^{-11}$~mbar directly before being loaded into the STM head. 

Clean NbSe$_2$ surfaces were obtained by cleaving NbSe$_2$ single crystal and (LaSe)$_{1.14}$(NbSe$_2$)$_2$ in ultrahigh vacuum at room temperature. The K-doped surface has been obtained by evaporating K atoms \textit{in-situ} from a carefully outgassed SAES getter source onto the NbSe$_2$ surface kept at $\sim$50 K to inhibit K intercalation. During the K evaporation, the pressure was maintained below 5 $\times$ 10$^{-10}$mBar. The ARPES measurements were carried out using a Scienta DA30 photoelectron analyser with He-I radiation ($h\nu$=$21.22$ eV). The total energy resolution was 5 meV and the base pressure during experiments was better than 1.5 $\times$ 10$^{-10}$ mbar.

Density Functional Theory (DFT) calculations were performed by using the \textsc{Quantum-Espresso} (QE) package\cite{QE1,QE2}. 
We adopted the generalized gradient approximation in the PBE parametrization\cite{PhysRevLett.77.3865} for the exchange-correlation functional.
The kinetic energy cutoff for plane-wave expansion has been set to 45 Ry (537 Ry for the charge density). In the case of non-collinear spin orbit coupling (SOC) calculation, we used ultra-soft fully-relativistic pseudopotential, directly downloaded from the QE repository\cite{QEpseudi}. In this case we used an energy cutoff of 75 Ry (537 Ry for the charge density).
Integration over the the Brillouin Zone (BZ) was performed using an uniform 4$\times$1$\times$1 Monkhorst and Pack\cite{Kgrid} grid.
For all of the calculation we adopt a Gaussian smearing of 0.01 Ry. For more details see supporting information.

The structural optimization for the bulk has been carried out by using as starting guess the experimental work of Roesky {\it et al.}\cite{LaNbSestructure}. We built a $\sqrt{3}\times 7$ NbSe$_2$ bilayer supercell of the compressed hexagonal unit cell and piled it up to a $1\times4$ supercell of the LaSe bilayer (see SI for more details). Similar calculations have been carried out for the NbSe$_2$ terminated surface. 

QPI spectra have been calculated adopting the Green function formalism, by using the following expression $\rho({\bf q },\omega)=-\frac{1}{\pi N_{\bf k}}{\rm Im}\sum_{n{\bf k},m} M_{n{\bf k},m{\bf k}+{\bf q}}G^0_{n{\bf k}}(\omega)G^0_{m{\bf k}+{\bf q}}(\omega)$
where $G^0_{n{\bf k}}(\omega)=1/(\epsilon_{{\bf k}n}-\omega-i\eta)$ is the Green function obtained from the relativistic band structure $\epsilon_{{\bf k}n}$ and $M_{n{\bf k},m{\bf k'}}$ is the matrix element due to scalar impurities that guarantees the spin conservation in the scattering process. As QPI spectra require ultradense
k-point grids we use the Wannier90 code\cite{MarzariComposite,MarzariEntangled,Pizzi2020} and perform the integration over an uniform $120\times 120$ k-points grid.

\vskip 0.5cm
{\bf Data availability.}
The datasets generated and analyzed during the current study are available from the corresponding author (T. C.) upon reasonable request.

\vskip 0.5cm
{\bf Acknowledgments.}
This work was supported by the French Agence Nationale de la Recherche through the contract ANR 15-CE30-0026-02 and ANR-19-CE24-0028. M. Calandra and M. Campetella acknowledge support from the Graphene Flagship (core 3). Calculations were performed at project RA4956 and on the Jean Zay supercomputer (Grand Challenge Jean Zay). T.Samuely and O.\v{S}ofranko were supported by VEGA 1/0743/19, APVV-18-0358 and SK-FR-2017-0015, COST action CA16218 Nanocohybri, and H2020 Infraia 824109 European Microkelvin Platform. C.Monney acknowledges funding by the Swiss National Science Foundation (SNSF), Grant No. P00P2$\_$170597. C. Monney, T. Jaouen and G. Kremer are very grateful to P. Aebi for sharing his ARPES setup.

\vskip 0.5cm
{\bf Author contributions.}
S. Sasaki and L. Cario grew the crystals and analysed them by X-ray diffraction and X-Ray spectroscopy. R.T. Leriche, A. Palacio-Morales, C. Brun,  I. Arfaoui, O. Šofranko, T. Samuely and T. Cren contributed to STM/STS and transport experiments. F. Debontridder and P. David supplied technical support for the STM setups. T. Jaouen, C. Monney and G. Kremer performed the ARPES experiments. M. Campetella, C. Tresca and M. Calandra performed the DFT calculations.

\vskip 0.5 cm
{\bf Competing interests.}
The authors declare no competing interests.

\bibliography{LaNb2Se5}

\end{document}